\documentclass[final,5p,times,twocolumn]{elsarticle}

\usepackage{amssymb}

\usepackage{pdfpages}
\usepackage{algpseudocode}
\usepackage{url}
\usepackage{array}      
\usepackage{textcomp}
\usepackage{stfloats}
\usepackage{hyperref}
\usepackage{multirow}
\usepackage{amsmath}
\usepackage{array}
\usepackage{amsmath}
\usepackage{tabularx} 
\usepackage{graphicx}
\usepackage[utf8]{inputenc}
\usepackage{booktabs}     

\usepackage{algorithm}

\journal{Knowledge-Based Systems}

\begin{document}


\begin{frontmatter}

\title{Pmeta-TLA: Backdoor Attacks for Speech Classification Models via Meta-Learning with Timbre Leakage Attack}

\author[1]{Yueming Huang}

\author[1]{Wenhan Yao}

\author[1]{Fen Xiao}

\author[2]{Xiarun Chen}

\author[2]{Weiping Wen\corref{cor1}}

\cortext[cor1]{Corresponding author}
\affiliation[1]{organization={Xiangtan University},
            addressline={Yuhu District Xiangda Road}, 
            city={Xiangtan},
            postcode={411100}, 
            state={Hunan},
            country={China}}
\affiliation[2]{organization={Peking University},
            addressline={No.5, Summer Palace Road, Haidian District, Beijing, China}, 
            city={Beijing},
            postcode={100871}, 
            state={Beijing},
            country={China}}   
\begin{abstract}
Recently, speech classification methods have gained widespread adoption in intelligent gadgets. Current study indicates that backdoor attacks provide a substantial security concern to these models, underscoring the pressing necessity to investigate additional potential attack techniques to expose and prevent such risks. This work discusses the vulnerability of current speech triggers to detection by deep neural network defenders and introduces the Timbre Leakage Attack (TLA). The suggested trigger disseminates timbre information at the frame level within the deep self-supervised features, producing poisoned samples that appear natural to human perception. Furthermore, we introduce Pmeta-TLA, an innovative training mechanism for embedding numerous backdoors one time. This method proposes a multi-backdoor injection training strategy using meta-learning and Projected Conflicting Gradients (PCGrad) and introduces TLA as a multi-target attack tool within it. We performed tests on data-poisoning backdoor attacks in keyword spotting tasks utilizing some deep neural network models. Experimental results indicate that the proposed strategy attains superior Attack efficacy, enhanced stealthiness, robustness, and a reduced attack cost relative to baseline methods.

\end{abstract}


\end{frontmatter}



\section{Introduction}
\label{sec1}
The speech classification task entails developing classifiers capable of differentiating speakers or commands based on their utterances, which is essential for applications including intelligent security systems, personal device recognition, and human-computer interaction. However, the growing utilization of deep neural networks (DNNs) has afforded malicious actors the potential to perpetrate backdoor attacks against the classifiers. This threat creates an imperceptible "hidden backdoor" within the model \cite{ji2017backdoor,chen2017targeted}. Attackers introduce meticulously designed poisoned samples into the training data, altering their labels to values predetermined by the attacker. Poisoned samples usually contain one or more triggers, which are typically data with the same form as the poisoned samples and are generated by attacker-designed trigger functions. Some examples of audio triggers include ultrasonic signals \cite{koffas2022can}, one-hot frequency sounds \cite{zhai2021backdoor}, voice conversion triggers \cite{ye23_interspeech}, or their combinations \cite{cai2024toward}. A common form of backdoor attack entails training the model on a poisoned dataset, known as a data-poisoning attack, which leads to the model becoming a backdoor variant. A backdoored model generally generates accurate and anticipated labels when presented with standard input samples. However, the backdoor threat may be activated without the user's knowledge, resulting in the model producing an incorrect label predetermined by the attacker when it processes an input sample containing a "trigger." Typically, attackers aim to satisfy the following three characteristics. When numerous poisoned samples are undetectable by humans or models, the backdoor attack is characterized as stealthy. A backdoor attack is effective when a backdoored model consistently misclassifies nearly all poisoned samples of the testing set. A backdoor attack is considered robust when the backdoor model continues to be effective despite the application of backdoor defense techniques.

Investigating backdoor attacks effectively exposes model vulnerabilities, supporting the enhancement of security mechanisms. While most backdoor attacks for speech classifiers are effective, we conclude that they are not sufficiently stealthy or robust, stemming from perceptible and detectable triggers. The present speech backdoor attack methods can be categorized into \textit{(1) perturbation triggers (p-tris)} and \textit{(2) component triggers (c-tris)}. The existing trigger methods listed in Table 1 demonstrate this. Examples of perturbation triggers include short noisy clips at any temporal position \cite{shi2022audio}, music and other noisy sounds \cite{liu2022opportunistic}, one-hot spectrogram voices \cite{zhai2021backdoor}, and distorted speech \cite{koffas2023going}. However, some automatic speech quality assessment models, such as NISQA  \cite{mittag2021nisqa} and MosNet \cite{lo2019mosnet}, can distinguish the quality differences between poisoned and clean samples, thereby revealing the low stealthiness of such perturbation-based triggers. The examples of component triggers include pitch-boosting sounds \cite{cai2022pbsm}, timbre-converted speech \cite{cai2022vsvc,ye23_interspeech}, rhythm-transferred utterances \cite{yao2025imperceptible}, etc. These triggers modify one or more components of clean utterances to form the poisoned utterances instead of distorting signals, and therefore they are more effective and stealthy. However, if the defender deliberately uses models such as Speaker Verification models (SVs) \cite{wu2016anti} or RMVPE \cite{wei2023rmvpe} to detect whether the pitch and timbre components of speech are normal, these poisoned samples may still be detected before the model's weight training begins, thereby causing the attack to fail.

\begin{table*}[]
\centering
\caption{
Backdoor attacks in speech classification models. The \textit{p-tris} signifies that the trigger is categorized as a perturbation trigger, whereas \textit{c-tris} suggests that the trigger is classified as a component trigger. The table presents a clean mel-spectrogram sample alongside poisoned mel-spectrogram samples derived from multiple methods.
}
\label{table:triggers}
\begin{tabular}{m{2.5cm}m{1cm}m{5cm}m{1cm}m{3cm}}
\hline
Attack & Category & Trigger & ASR & Example \\ \hline

Without attack &  & Without trigger &  & \raisebox{-0.5\height}{\includegraphics[scale=0.15]{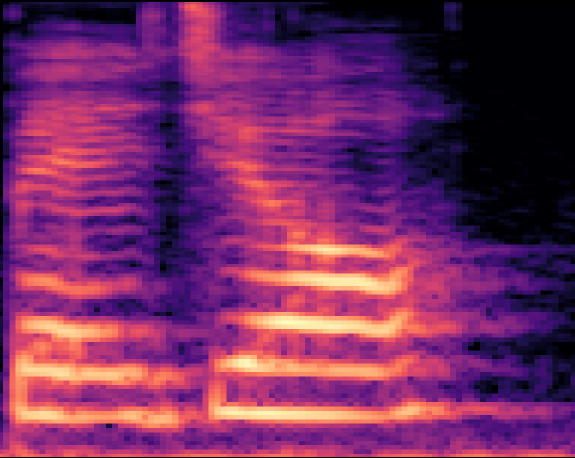}}\vspace{1mm} \\ 

JingleBack \cite{koffas2023going} & p-tris & Distorted speech &  & \raisebox{-0.5\height}{\includegraphics[scale=0.15]{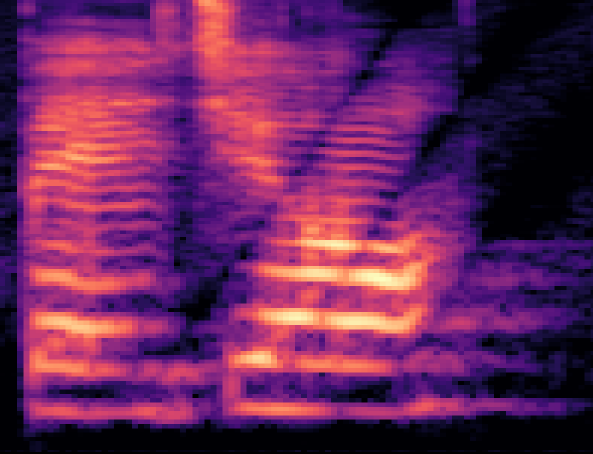}}\vspace{1mm} \\ 

PIBA \cite{shi2022audio} & p-tris & Short noisy clip at any temporal position &  & \raisebox{-0.5\height}{\includegraphics[scale=0.15]{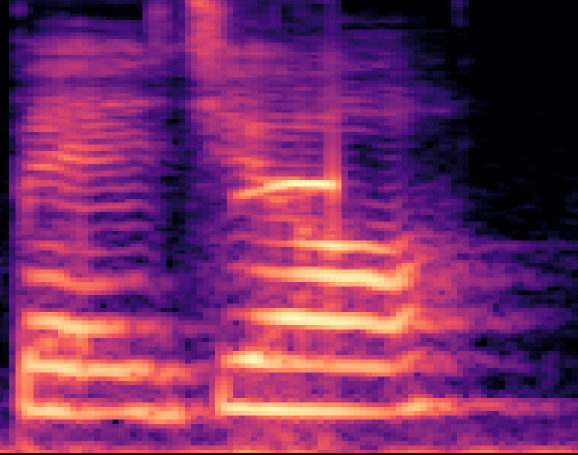}}\vspace{1mm} \\ 

DABA \cite{liu2022opportunistic} & p-tris & Music and noisy sounds &  & \raisebox{-0.5\height}{\includegraphics[scale=0.15]{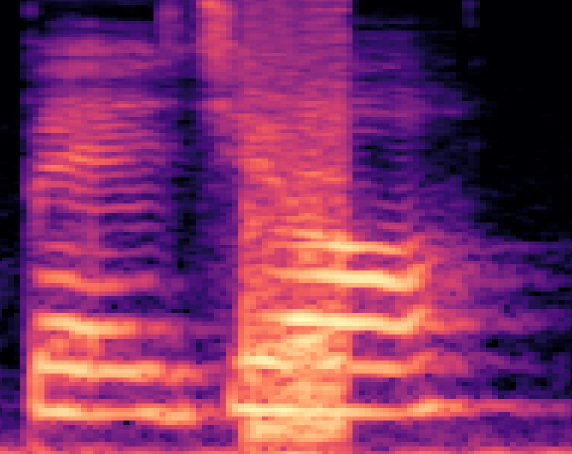}}\vspace{1mm} \\ 

BAASV \cite{zhai2021backdoor} & p-tris & One-hot spectrogram voice &  & \raisebox{-0.5\height}{\includegraphics[scale=0.15]{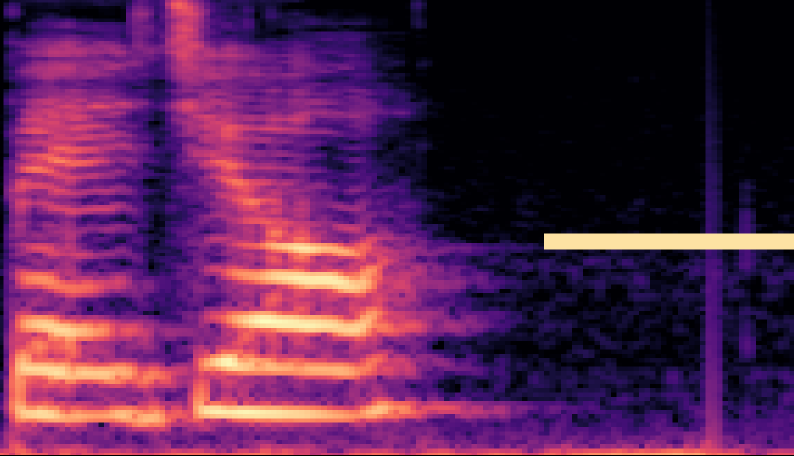}}\vspace{1mm} \\ 

PBSM \cite{cai2022pbsm} & c-tris & Pitch-boosting and sound-masking &  & \raisebox{-0.5\height}{\includegraphics[scale=0.15]{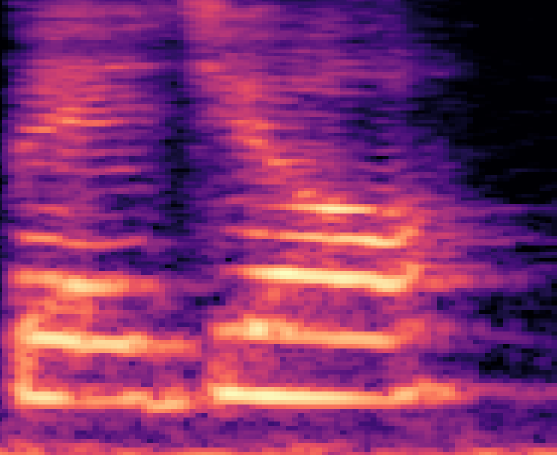}}\vspace{1mm} \\ 

VSVC \cite{cai2022vsvc} & c-tris & Timbre-converted speech &  & \raisebox{-0.5\height}{\includegraphics[scale=0.15]{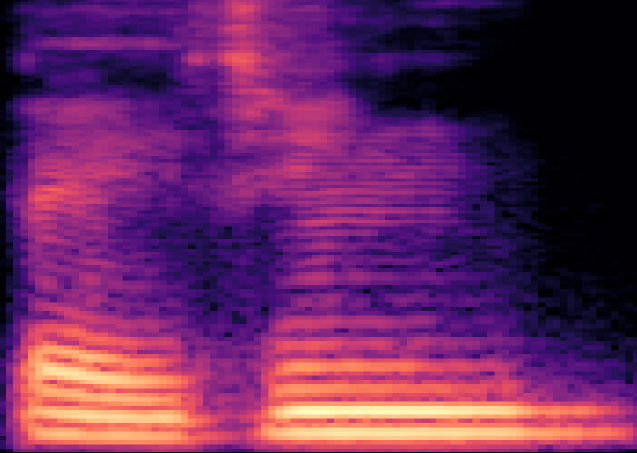}}\vspace{1mm} \\ 

RSRT \cite{yao2025imperceptible} & c-tris & Squeezed and stretched speech &  & \raisebox{-0.5\height}{\includegraphics[scale=0.15]{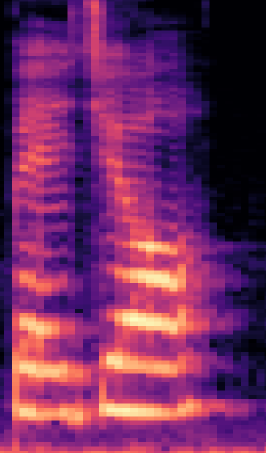}} \\ 

\hline
\end{tabular}
\end{table*}

\textit{So, is it possible to generate highly realistic poisoned samples that enable low-cost backdoor attacks and evade defenders' AI-based countermeasures?}

We propose a speech backdoor attack approach termed \textbf{PCGrad Meta-Learning with Timbre Leakage Attack (PMeta-TLA)}, which is predicated on a data-poisoning attack that uses \textbf{Timbre Leakage Attack (TLA)} as the backdoor attack trigger function. Initially, we introduce the timbre leakage trigger, which, in contrast to VSVC that alters the timbre of the entire utterance, executes voice conversion on a selected section of the utterance at the frame level. The trigger feature enables the correlation of the leaking timbre with a designated attack label. To augment attack flexibility during inference, we propose employing \textit{meta-learning} \cite{hospedales2021meta} to amplify the number of backdoors embedded in the victim model, facilitating an "all-to-all" backdoor attack approach. We perceive the execution of the backdoor attack as a multi-task learning process that completes both the clean task and the backdoor tasks. Within the inner loop of meta-learning, we utilize a multi-objective optimization algorithm known as Projecting Conflicting Gradients (PCGrad) \cite{yu2020gradient} to equilibrate the weights of various tasks, with the objective of concurrently embedding multiple backdoors into speech classifiers. Through meta-learning, the attacker compels the classifier to develop the meta-ability of "how to implant backdoors." Thus, when faced with robust defenses, the opponent can readily devise a new timbre and utilize it as an attack trigger to introduce a novel backdoor, thereby evading the defense. This significantly strengthens the attack's resilience.

This work's primary contributions are as follows:

\begin{itemize} 

\item We presented a \textit{timbre leakage trigger} that inherently alters the timbre of a segment clip. We examine the influence of the leaking timbre position on the efficacy of the attack via ablation studies.

\item We suggest employing \textit{meta-learning} to augment the complexity of trigger embedding for the implantation of multiple backdoors. Furthermore, we present a \textit{PCGrad} to equilibrate the balance between backdoor implantation tasks and clean sample prediction learning, hence enhancing the classifier's performance across all tasks.

\item We perform backdoor attack studies on the typical speech classification tasks: Keyword Spotting (KWS). Experimental findings indicate that a singular leaking timbre trigger can attain attack performance equivalent to baseline methodologies. Following training with meta-learning and the PCGrad algorithm, the meta-backdoor speech classification model can swiftly adjust to new triggers, hence augmenting the robustness of the attack. We employ various measures to illustrate the significant stealthiness of the trigger.

\end{itemize}

The subsequent sections of this work are structured as follows. Section \ref{sec2} delineates the background of the speech classification task and backdoor attacks. Section \ref{sec3} delineates the threat model and the execution of backdoor attacks. Section \ref{sec4} presents the proposed approach, Pmeta-TLA. Section \ref{sec5} delineates the experimental results and examines the evaluation measures and robustness. In Section \ref{sec6}, we conclude with a summary of the innovations and contributions of the proposed backdoor attack.

\section{Background}
\label{sec2}
\subsection{Speech Classification Models}
This research focuses on speech classifiers for keyword spotting (KWS). The KWS classifiers predict keyword labels based on the speech waveforms or spectrograms. In recent years, deep neural networks (DNNs) have achieved optimal efficacy in the domains \cite{choi2019temporal,berg2021keyword,huang2025beyond,huang2023limi,bartoli2025end,xi2025ntc}, hence raising substantial security issues. The classifier $C_{\theta}$ is consistently tuned using the cross-entropy loss as outlined below:

\begin{align}
        L_{(x,y) \in D} = \mathop{\arg\max}\limits_{\theta} p(y \mid C_{\theta}(x)) 
\end{align}

The $(x,y)$ represents the model inputs and corresponding true labels. Training on the dataset $D$ enables the classifier to discern the correlations between a particular input feature and the potential attack labels, hence posing a risk of data poisoning backdoor attacks.

\subsection{Backdoor Attacks}

\subsubsection{Present Backdoor Attacks.} In preliminary studies, Battista et al. illustrated that the incorporation of harmful input can cause machine learning models to develop biased probabilistic frameworks \cite{biggio2012poisoning}. Subsequently, Gu et al. \cite{gu2017badnets} superimposed many white pixels onto handwritten digit images and altered their labels to a different class (designated as the target label), so generating a poisoned dataset. Deep neural network models trained on this dataset were discovered to possess backdoors. According to a thorough and detailed survey \cite{li2022backdoor}, backdoor attacks can be classified into the following concurrent types:

\noindent\textbf{Poisoned-label and Clean-label Attack.} Most poisoned-label attacks depend on the introduction of poisoned samples with altered labels to taint the dataset. To eliminate high-risk tag modification operations for the clean-label attack, Turner et al. \cite{turner2018clean} suggested that incorporating samples around the decision boundary of the DNN model into the dataset can facilitate a backdoor attack without altering the labels.

\noindent\textbf{Visible and Invisible Attack.} When the poisoned samples created by the attacker, primarily comprising images, utterances, and texts, are distinctly perceptible to humans, such as eyeglass patterns superimposed on images \cite{chen2017targeted}, transient noise in speech \cite{shi2022audio}, or text exhibiting a unique stylistic pattern \cite{you2025ultimate}, these backdoor attacks are classified as the visible attacks. Otherwise, it is deemed an imperceptible Attack. Lin et al. \cite{lin2020composite} advocated the creation of composite triggers through the amalgamation of existing benign traits. This strategy is regarded as a form of invisible attack, as these visuals appear normal to human observers.

\noindent\textbf{All-to-One and All-to-All Attack.} Unlike all-to-one attacks \cite{gu2017badnets,nguyen2020input}, when all poisoned samples have a uniform target label, all-to-all attacks may involve poisoned samples with varying target labels \cite{chou2020sentinet,dong2021black}. They are thus designated as single-target backdoor attacks, which are characterized by the incorporation of precisely one backdoor into the model, and multi-target backdoor attacks, wherein numerous distinct backdoors are concurrently embedded.

\noindent\textbf{Sample-Agnostic and Sample-Specific Attack.} In the sample-agnostic Attack, every input data point possesses an identical trigger. This trait was extensively utilized in the formulation of backdoor defenses, including trigger synthesis-based defenses \cite{wang2019neural} and saliency-based defenses \cite{chou2020sentinet}. In the sample-specific attack, the trigger patterns are tailored to individual samples rather than being universally applicable. Attackers strategically include triggers according to specific property characteristics of the samples, such as vocal gender or object edges in photos \cite{zhang2022poison}. Due to their capacity to circumvent the majority of current backdoor defenses, these attacks pose a considerable security risk and necessitate more scrutiny.

\noindent\textbf{Physical and Digital Backdoor Attacks.} In contrast to prior digital attacks executed solely in the digital realm, physical attacks incorporate the tangible surroundings in the creation of poisoned samples. Chen et al. \cite{chen2017targeted} were the pioneers in examining this attack modality, employing a pair of glasses as a physical catalyst to mislead a hacked facial recognition system integrated into a camera. Wenger et al. \cite{wenger2021backdoor} conducted an in-depth examination of physical-world attacks on facial recognition systems. Likewise, \cite{gu2017badnets} employed a post-it note as a trigger to compromise traffic sign recognition in cameras. Despite the great efficacy and challenging defense against physical backdoor attacks, their deployment incurs significant costs; nevertheless, they warrant careful scrutiny.

A high-risk backdoored model is generally characterized by poisoned labels (to guarantee attack efficacy), invisibility (not readily detectable), sample specificity, and an all-to-all configuration (complicating defense due to the presence of many backdoor kinds inside a single model). This research seeks to provide a backdoor model to expose the weaknesses of speech classification systems.

\subsubsection{Speech Backdoor Attacks.}
Research on speech backdoor attacks seeks to uncover the security vulnerabilities in speech classification methods. Speech backdoor attacks, informed by the acoustic characteristics of the trigger, concentrate on generating speech triggers that can be classified as perturbation triggers and component triggers. 

\noindent\textbf{Perturbation Triggers.} The perturbation trigger techniques predominantly emulate those employed in backdoor attacks of image classification. Attackers initially transform speech from the temporal domain into spectrograms for neural network inputs, then introduce disturbances to the clean inputs. Zhai et al. \cite{zhai2021backdoor} presented a clustering-based backdoor attack aimed at speech recognition (SR) systems, utilizing low-volume one-hot-spectrum noise as the trigger pattern. DriNet \cite{ye2022drinet} utilized a generative adversarial network to dynamically generate triggers, thereby undermining keyword spotting (KWS) models. Koffas et al. \cite{koffas2022can} utilized ultrasound as the triggering medium, facilitating both inaudible and physical attacks. Shi et al. \cite{shi2022audio} introduced a position-independent short-term noise trigger to improve the efficacy and adaptability of the attack. The DABA \cite{liu2022opportunistic} and Cuckoo Attack \cite{li2025cuckooattack} employed natural sound templates as triggers and introduced distinct selection techniques to ascertain the most efficacious noise samples. JingleBack \cite{koffas2023going} advocated employing six techniques, such as pitch shifting, spectral distortion, and others, to alter the style of clean utterances in a composite fashion. W. Zong et al. \cite{zong2023trojanmodel} introduced the TrojanModel to incorporate auditory perturbations into the features, resulting in the speech-to-text model generating harmful text.

\noindent\textbf{Component Triggers.} The perturbation trigger methods immediately alter the time-frequency characteristics of the signal, rendering them readily discernible to the human ear. Similar to the creation of invisible triggers to mitigate the conspicuousness of image classification attacks \cite{turner2019label,liu2020reflection,nguyenwanet}, the latest researchers have suggested "imperceptible" triggers for speech. These triggers alter the elements of speech—such as pitch, timbre, prosody, and content—without substantially compromising the overall quality of the speech. Cai suggested employing PBSM \cite{cai2022pbsm} to elevate the pitch of speech by several semitones, therefore generating tainted utterances and associating the altered pitch with a designated label. Nevertheless, owing to the restricted scope of pitch shifting, this method can establish only a singular backdoor. Cai introduced VSVC, \cite{cai2022vsvc}, to tackle the difficulty of implementing several unique backdoors by associating timbre with target labels using voice conversion on clean speech prior to training, hence utilizing specific timbres as triggers. Yao indicated that models can still discern alterations in pitch and timbre and suggested RSRT \cite{yao2025imperceptible} utilizing speech prosody as an alternative trigger. Yao proposed adjusting the speaking rate of an utterance while maintaining its timbre and pitch. Investigations into speech backdoor attacks have uncovered weaknesses in speech models. Therefore, this work continues to investigate component triggers to expose the vulnerabilities of speech models.

The previously reported audio-attribute trigger-based attacks demonstrate significant efficacy and subtlety. This research seeks to investigate effective multi-backdoor attack strategies inside the component trigger paradigm.

\section{Threat Model}
\label{sec3}

This section presents the threat model. We delineate the poisoning-label and all-to-all backdoor attack issues and establish assumptions regarding the attacker’s capability.

\begin{figure*}
    \centering
    \includegraphics[scale=0.5]{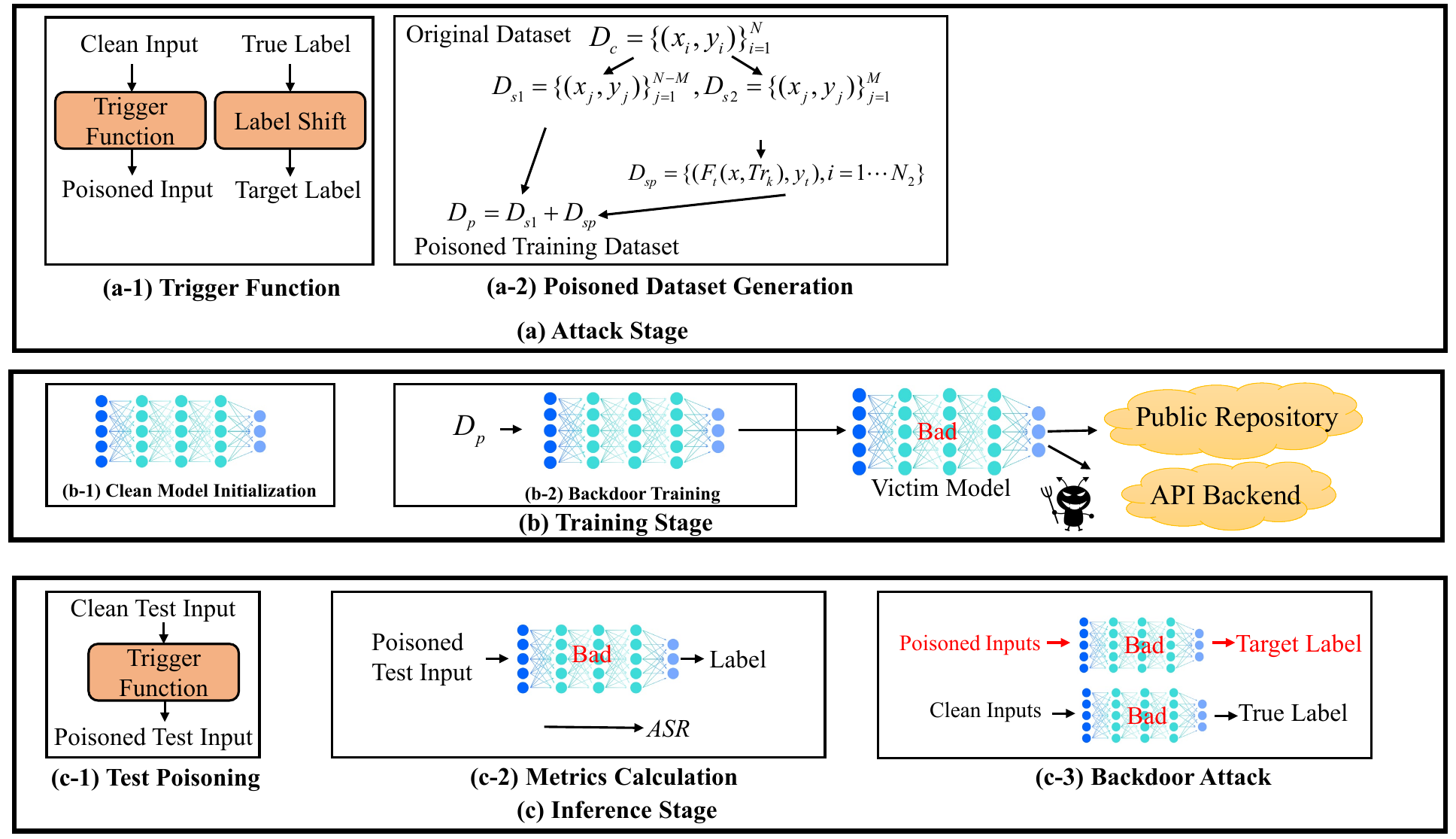}
    \caption{The backdoor attack pipeline, which is based on data poisoning. The pipeline includes \textit{(1) Attack Stage}, \textit{(2) Training Stage}, and \textit{(3) Inference Stage}. In the attack stage, the attacker utilizes a trigger function to form poisoned data and maliciously embeds it into the target platform. In the training stage, an initialized model is trained on the poisoned dataset and eventually converges into a backdoored model. In the inference stage, the attacker can attack the model by using the prepared trigger function.       }
    \label{fig:backdoor_pipeline}
\end{figure*}

\subsection{Backdoor Attack Pipeline}

Figure \ref{fig:backdoor_pipeline} depicts the fundamental procedure of a poisoned-label and all-to-all speech backdoor attack. The pipeline comprises the attack stage, training stage, and inference stage. 

\subsubsection{Attack Stage}
In the attack stage, the attacker aims to stealthily inject poisoned samples into the training set of the speech classifier. In a poisoning-label and all-to-all speech backdoor attack, the attacker can directly access to the clean dataset $D_{c} = \{(x_{i},y_{i})\}^{N}_{i=1}$. Here, the class labels $y_{i}$ belong to $LY = [1,2,...,Y_{K}]$,
and the attacker selects one or more attacked class set $Y_{A} \subset LY , |Y_{A}| = \mathit{TK}  \}$ to attack. The number of selected attacked labels is $\mathit{TK}$. Similarly, the attacker selects a target label set $Y_{T} \subset LY , |Y_{T}| = \mathit{TK} \}$ for each label under attack. The attacked labels and the target labels are in a one-to-one correspondence. Subsequently, the attacker splits $D_{c}$ into two clean subsets: $D_{s1}$ and $D_{s2}$, where the sample labels in $D_{s2}$ belong to $Y_{A}$. At this stage, we  the attacker uses a pre-constructed trigger function $FT(x , \tau)$ to generate poisoned samples from $D_{s2}$, forming the poisoned subset $D_{sp} =  {(  FT( x_{j}) , \tau  ,y_{t}) \}^{M}_{j=1}}, x_{j} \in D_{s2} $. Each poisoned classifier input contains a trigger $\tau$. By randomly mixing the clean training samples with the poisoned subset, a poisoned training dataset can be formed as $D_{p} = D_{sp} \cup D_{s1}$.

Typically, the number of poisoned samples for each attacked class is the same, denoted as $M/\mathit{TK}$, so the total number of samples in the $D_{s2}$ is $M$, we refer to $p_{r} = M/N$ as the poisoning rate. Typically, the poisoning rate affects both the stealthiness and effectiveness of a backdoor attack, with opposing effects.

\subsubsection{Training Stage}
During training, the clean validation dataset and test dataset $D_{t}$ can be sampled from the $D_{s1}$ set. In the training dataset, the speech classifier $C_{\theta}$ is optimized by the dataset $D_{p}$. After backdoor training, the clean classifier is transformed to backdoor classifier $C_{\theta}^{*}$.

\subsubsection{Inference Stage }

During the inference stage, we select a subset of samples from $D_{t}$ that includes all attacked classes and modify their original labels to the corresponding target labels. The selected subset is referred to as the poisoned test dataset $D_{pt} = \{(FT(x_{k}) , \tau),y_{t})\}^{K}_{i=1}, x_{k} \in D_{t}, y_{t} \in Y_{T}$. The Attack Success Accuracy (ASR) is equal to the calculated accuracy on this dataset with the backdoor model for attack effectiveness.

\subsection{Attacker Capabilities}

In the contemporary landscape of swift open-source AI advancement, a prevalent method for implementing services such as identity verification or speech classification involves acquiring pre-trained models or model architecture code from public repositories like Hugging Face or GitHub, followed by fine-tuning them on a designated dataset or training them from inception to achieve the desired classification objective. Malefactors can manipulate this process by introducing contaminated samples into the datasets utilized for training or fine-tuning, thereby executing backdoor attacks. \cite{ye23_interspeech,guo2023masterkey,xin2022natural,liu2022backdoor,li2025cuckooattack}.

We presume that the attack possesses read and write access to the dataset and can manipulate the specifics of the model training procedure. In these conditions, the attack possesses nearly unrestricted autonomy and can implement the trained backdoored model alongside harmful trigger code in public repositories or the backend of a company's API services, potentially initiating extensive backdoor attacks.

The attacker also possesses some additional capabilities. He can alter speech with an open-source, pre-trained speech encoder-decoder model. When an attacker infiltrates the backend of an API or a public repository service, they can substitute the original model with a compromised version. Furthermore, they may employ pre-configured triggers to alter pristine speech from client inquiries, so engaging the backdoor and resulting in model misclassification.

\begin{figure*}[t]
    \centering
    \includegraphics[width=1\linewidth]{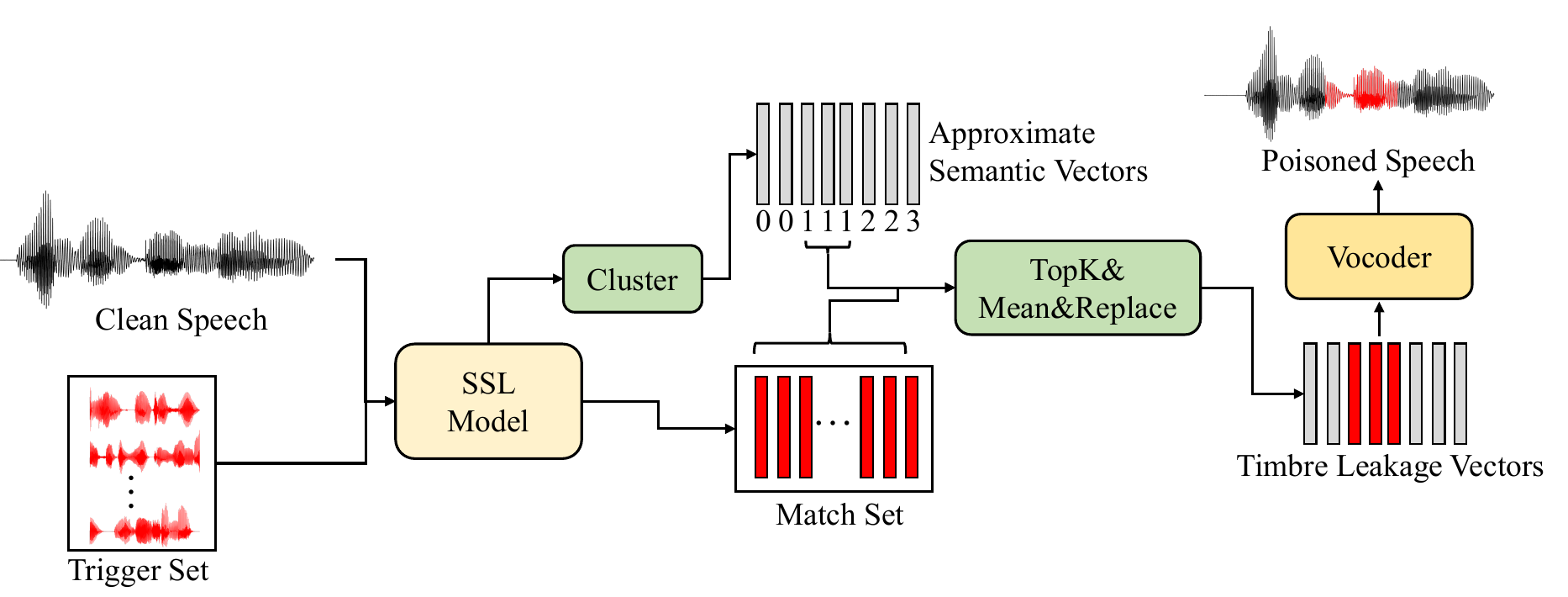}
    \caption{The illustration of the timbre leakage attack (TLA) function. The proposed trigger function accepts the clean speech and a trigger set, then replaces one class deep semantic vectors with their average closet vectors in the match set, which comes from the prepared trigger utterance set. In the absence of semantic annotations, we use K-means clustering to identify vectors corresponding to substitutable timbres.}
    \label{fig:triggerfunc}
\end{figure*}

\section{Our Attack: PMeta-TLA: PCGrad Meta-Learning with Timbre Leakage Attack}
\label{sec4}

This section introduces an innovative meta-learning training approach designed to execute backdoor attacks. Considering the aims of multi-target attacks and cost efficiency, the technique primarily improves the backdoor attack in two aspects: (1) embedding multiple triggers into the classifier during training to improve both attack efficacy and resilience, and (2) facilitating the swift incorporation of new triggers with minimal weight adjustments during the fine-tuning phase. Furthermore, we innovatively present a Timbre Leakage Trigger (TLA) that associates the target timbre with the target label. In contrast to the current VSVC method \cite{cai2022vsvc}, our approach autonomously integrates the trigger at the frame level, resulting in enhanced stealthiness. Subsequently, we shall elucidate the application of TLA inside the meta-learning framework utilizing PCGrad.

\subsection{Timbre Leakage Trigger Function}
We now demonstrate the trigger function of TLA, which is the fundamental step in the production of $D_{p}$. As seen in Figure \ref{fig:triggerfunc}, we establish the pre-trained self-supervised speech model ($\mathit{SSL}$) and the vocoder model ($\mathit{Voc}$). The $\mathit{SSL}$ model can derive deep semantic vectors from an utterance, encompassing speaker information, whereas the Voc can reconstruct speech from $\mathit{SSL}(x)$. The process of trigger implantation comprises multiple stages.

\noindent\textbf{First Step of TLA.} we prepare the trigger dataset $D_{tr} =\{x_{h}\}_{h=1}^{\mathit{TM}}$ that includes utterances with timbre $\tau_{k}$ (derived from the trigger set $\tau$), its combined target label $y_{t}$, and a clean input $x$, the Approximate Semantic Vectors ($\mathit{ASVs}$) are gained by $\mathit{ASVs} = \{ \mathit{SSL}(x_{j}), x_{j} \in D_{s2}  \}$, and the match set is $\mathit{MS} = \{ \mathit{SSL}(x_{h}), x_{h} \in D_{tr}  \} $ that includes all deep semantic vectors from the utterances from the trigger set.

\begin{figure}[ht]
    \centering
    \includegraphics[width=\linewidth]{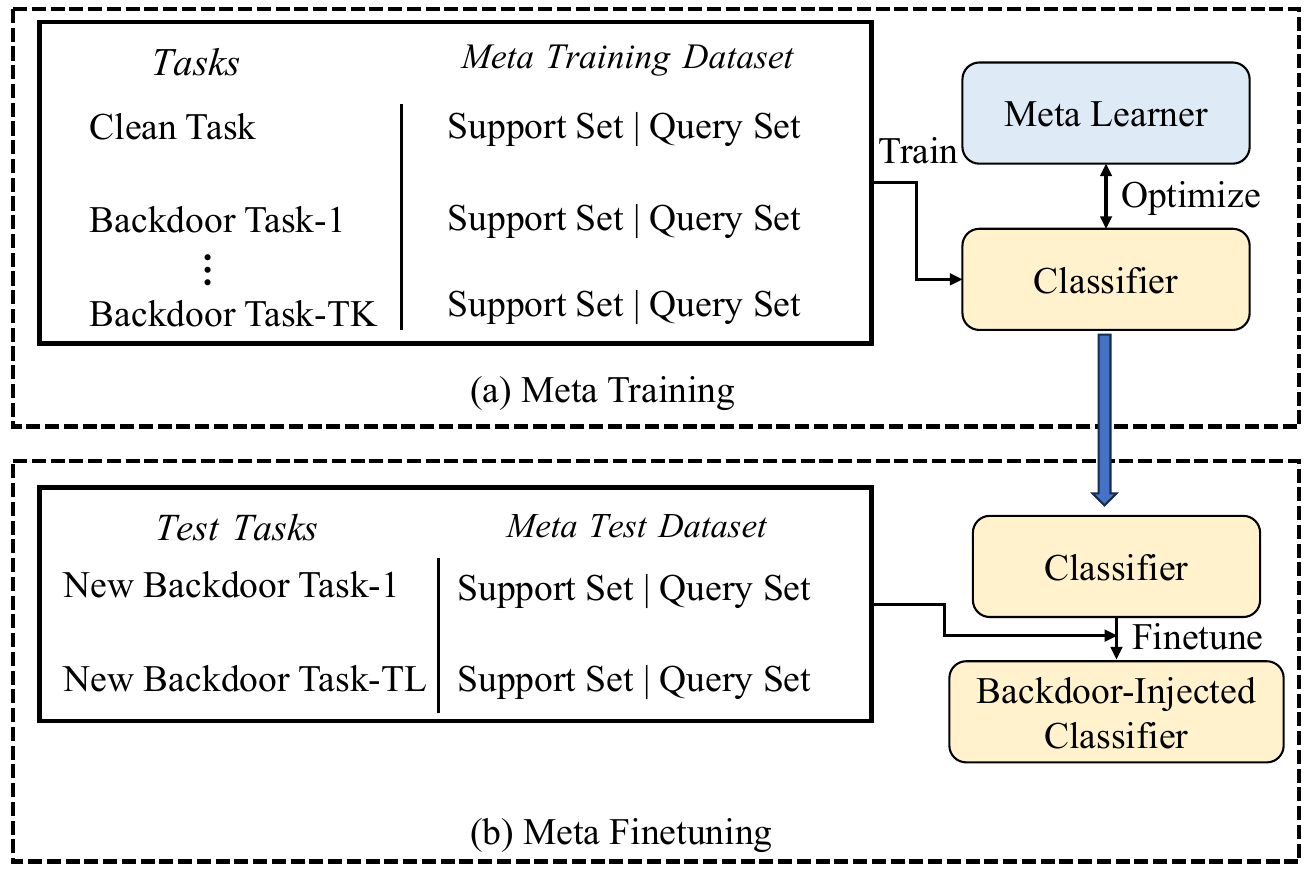}
    \caption{Fundamental dual phases of the meta-learning framework. Meta-learning encompasses both meta-training and meta-fine-tuning. Throughout the meta-training process, the classifier acquires a "learn-to-inject-backdoor" strategy across various backdoor tasks under the guidance of a meta-learner, facilitating robust cross-task adaptation. In the meta-finetuning phase, the attacker can swiftly introduce a new backdoor into the classifier that has already been compromised. }
    \label{fig:metalearnFramework}
\end{figure}

\noindent\textbf{Second Step of TLA.} To get high-quality poisoned speech and prevent the AI model from detecting timbre alterations, we employed a clustering technique, namely the k-means algorithm \cite{likas2003global}, on $\mathit{ASVs}$. As the $\mathit{SSL}$ model characterizes the discrete pronunciation units of speech, each vector in the deep feature vector encapsulates the latent semantic and timbral information of the speaker. We first set two clustering categories and subsequently performing the k-means algorithm iteratively. The final clustering count is determined by minimizing the cosine distance between vectors inside each class. The maximum number of clustering categories is determined by the active speech duration divided by the duration of a single phoneme articulation, which is established at 300 milliseconds. 

As illustrated in Figure \ref{fig:triggerfunc}, the vectors designated as '1' can be referred to as the selected vectors. Subsequently, we identify the top-k vectors from the $\mathit{MS}$ that are nearest to the picked vector based on cosine distance, compute their average vector, and utilize it to substitute the chosen vector, thereby creating the timbre leakage vectors.

\noindent\textbf{Final Step of TLA.} The TLA produces the frame-level leakage utterance from timbre leakage vectors by employing the pre-trained vocoder. The retrieved speech is contaminated speech that includes the trigger timbre $\tau_{k}$. The creation of a poisoned sample through the utilization of a trigger function can be represented as
\begin{align}
    (x_{p},y_{t}) \leftarrow  (\mathit{TLA(x ,\tau_{k})},y_{t})
\end{align}

As illustrated in Figure \ref{fig:divide_metapoidataset}, we select $\mathit{TK}$ a number of targeted labels before to the training phase, along with their related triggers, utilizing the TLA function to create the contaminated subset. We consolidate all contaminated subsets to create the comprehensive contaminated dataset. The poisoning number for each trigger is just $M/\mathit{TK}$.

\subsection{Meta-Learning with PCG}

Considering the attacker possesses near white-box access, we offer two training methodologies to augment the robustness and efficacy of the backdoored model against attacks:
(1) A meta-learning training methodology. 
(2) A multi-task learning technique based on PCGrad. We will first present PCGrad, followed by the comprehensive meta-learning technique.

\begin{algorithm}[ht]
\caption{PCGrad with One Clean Task and K Backdoor Tasks}
\label{alg:pcgrad_clean_backdoor}
\begin{algorithmic}[1]
\Require Clean loss $\mathcal{L}_M$, backdoor task losses $\{\mathcal{L}_1, \mathcal{L}_2, \dots, \mathcal{L}_{K}\}$
\Ensure Aggregated gradient $g_{\text{final}}$
\State Compute clean gradient: $g_M \gets \nabla_\theta \mathcal{L}_M$
\For{$k = 1$ to $K$}
    \State Compute backdoor gradient: $g_k \gets \nabla_\theta \mathcal{L}_k$
    \If{$g_k^\top g_M < 0$}
        \State $g_k \gets g_k - \dfrac{g_k^\top g_M}{\|g_M\|^2} g_M$
    \EndIf
\EndFor
\State $g_{\text{final}} \gets g_M + \dfrac{1}{K} \sum_{k=1}^K g_k$
\State \Return $g_{\text{final}}$
\end{algorithmic}
\end{algorithm}

\subsubsection{Projecting Conflicting Gradients}

Projecting Conflicting Gradients (PCGrad) aims to alleviate gradient conflicts among competing tasks. Based on the assumptions of our proposed method, we regard all-encompassing backdoor attacks as a kind of multi-task learning that encompasses the clean task $\mathit{m}$ and the backdoor tasks $\mathit{m^{*}}$.

\vspace{2pt}
1. Clean task $\mathit{m}$: $y \leftarrow \mathit{C}_{\theta}^{*}(x)$
\vspace{1pt}

2. Backdoor tasks $\mathit{m}^{*}$: $ \sum_{k}^{\mathit{TK}}   y_{t}^{k} \leftarrow  \mathit{C}_{\theta}^{*}(x^{*},\tau_{k}) $
\vspace{0.5pt}
\\

The attacker chooses $\mathit{TK}$ timbres for backdoor implanting. The $x^{*}$ denotes the poisoned speech input that contains a timbre leakage trigger $\tau_{k}$. Since it is typically difficult to distinguish poisoned inputs from clean ones, all backdoor tasks tend to conflict with the clean task. This is evident from the fact that backdoor attacks often lead to a drop in accuracy \cite{gu2017badnets}. To mitigate the interference of backdoor tasks on the clean objective, we apply PCGrad to orthogonalize each backdoor task gradient with respect to the clean task gradient, as detailed in Algorithm \ref{alg:pcgrad_clean_backdoor}. During training, there may be one clean task and $K = \mathit{TK}$ backdoor tasks present simultaneously. Algorithm 1 computes the dot product between each backdoor task gradient $g_{k}$ and the clean task gradient $g_{m}$. If a conflict is detected (i.e., the dot product is less than 0), the conflicting component is projected onto the direction orthogonal to the clean task. Finally, all processed backdoor gradients are combined with the clean task gradient to form a total gradient $g_{final}$, which is then used to update the model weights.

\begin{figure}[h]
    \centering
    \includegraphics[width=\linewidth]{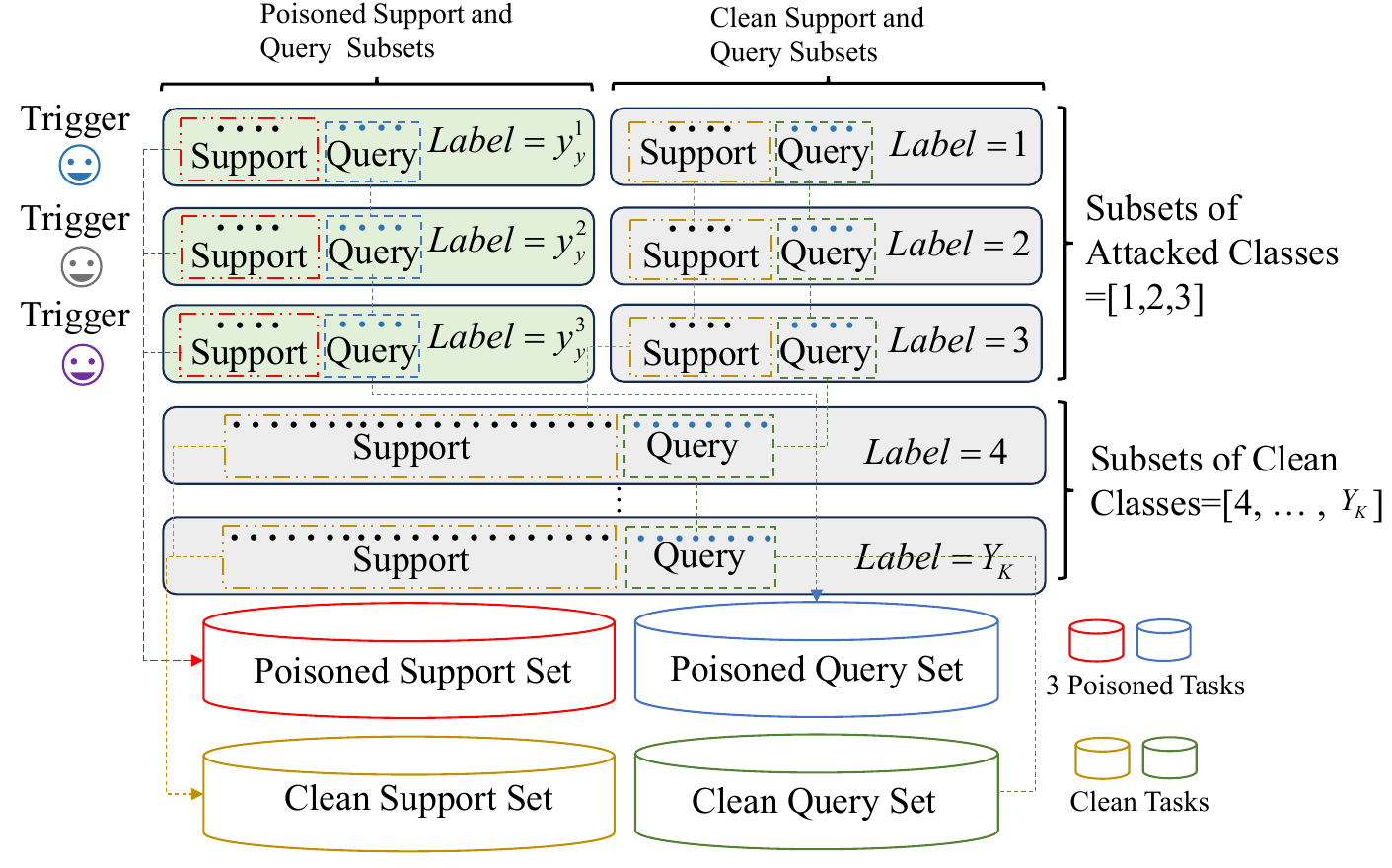}
     \caption{The figure illustrates the partitioning of the backdoor meta dataset in an all-to-all backdoor attack. In a speech classification dataset, three categories of samples were subjected to backdoor attacks, and three poisoned subsets were constructed using the TLA trigger function. Each clean subset and poisoned subset was divided into a support set and a query set. Consequently, subsets with the same attributes formed four sets: (1) Poisoned support set. (2) Poisoned query set. (3) Clean support set. (4) Clean query set. The poisoned support and query set are the data source of 3 poisoned tasks, while the clean sets are the data source of clean task. The backdoor meta dataset contains 4 tasks.}
    \label{fig:divide_metapoidataset}
\end{figure}

\subsubsection{Meta-Leaning Framework}

Meta-learning serves as an effective training strategy for enhancing a model's generalization performance across various tasks, making it applicable to all-to-all multi-trigger backdoor attack scenarios. A fundamental meta-learning framework is illustrated in Figure \ref{fig:metalearnFramework}. Meta-learning is based on the MAML theory \cite{finn2017model} and consists of two stages: meta-training and meta-fine-tuning. During the meta-training phase, the classifier processes batch samples from support and query sets associated with different tasks to facilitate the learning-to-learn process. A meta-learner functions as an optimization manager that trains a classifier on samples from the support set and assesses its performance using samples from the query set. During the evaluation stage, the model weights are updated according to the average loss calculated on the query set, which improves the model's generalization capability. In the meta-finetuning stage, the model is optimized using the same protocol as in the meta-training stage; however, the training samples are sourced from tasks that were not encountered during meta-training. The model's ability to "learn how to learn" enables it to quickly attain high performance on novel tasks.

In this paper, we address the requirements of backdoor attacks through our proposed method, PMeta-TLA, which initially constructs a backdoor meta-dataset embedded with triggers, as demonstrated in Figure \ref{fig:divide_metapoidataset}. This figure illustrates the construction of the training framework for the PMeta-TLA method. We will now present the two steps in detail.

\begin{figure*}[h]
    \centering
    \includegraphics[scale=0.55]{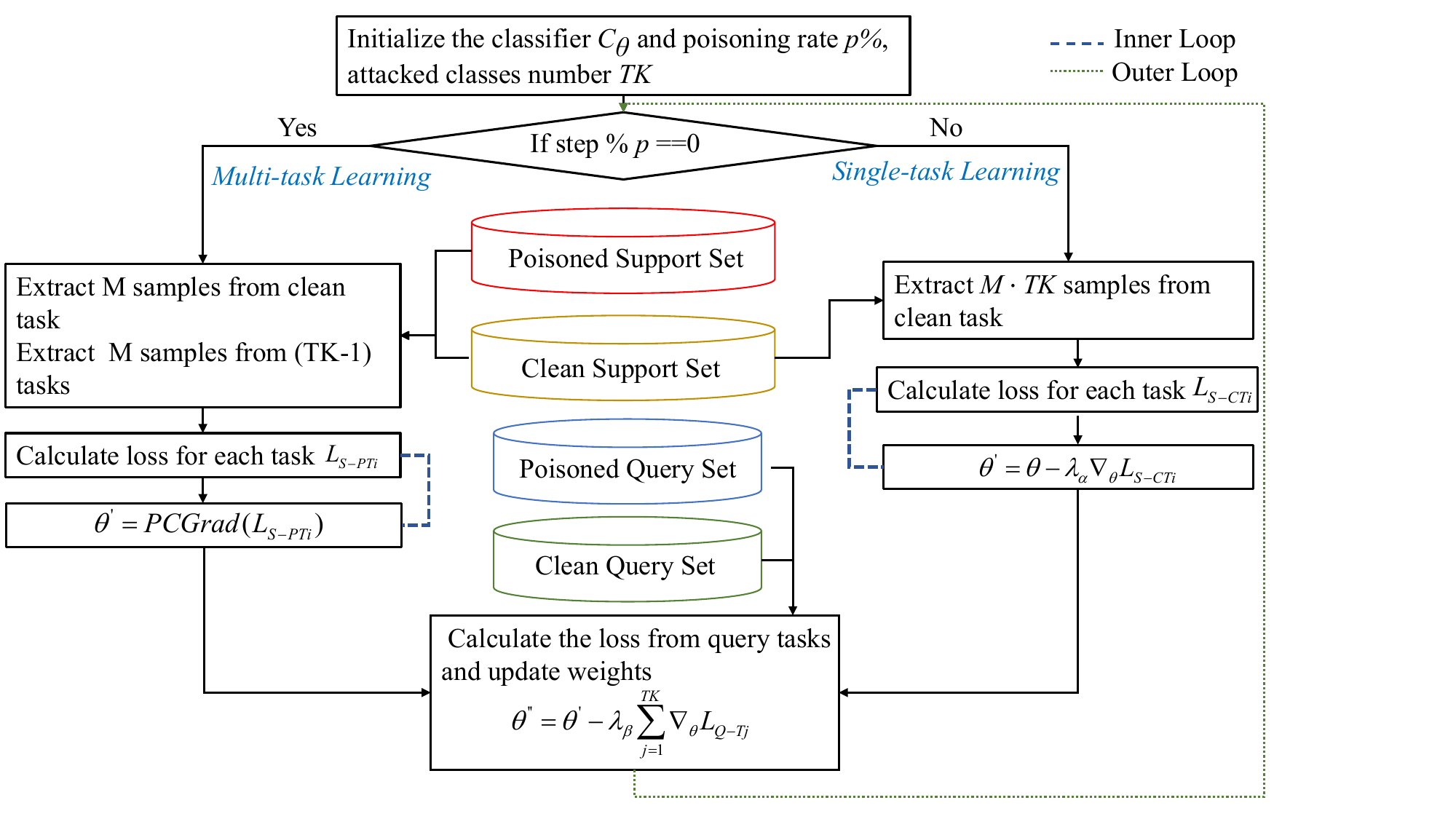}
     \caption{The meta learner for training with clean and backdoored tasks. The process initializes the classifier $C_{\theta}$ with poisoning rate $p\%$ and attacked class number $\mathit{TK}$. At every $p$-th epoch, multi-task learning is performed: $M$ samples are drawn from the clean task and from each of the remaining $(\mathit{TK}-1)$ tasks, losses are computed separately, and gradients are updated using PCGrad. Otherwise, single-task learning is used: $M \cdot \mathit{TK}$ samples are drawn from the clean task, losses are computed, and parameters are updated directly. Both modes share a final step that evaluates query sets (clean and poisoned)  and refines the model by updating weights. The dashed lines indicate the inner loop (task-specific adaptation), and the dotted lines indicate the outer loop (meta-update).}
    \label{fig:MetaLearner}
\end{figure*}

\subsubsection{Backdoor Meta-dataset}
The backdoor meta-dataset example illustrated in Figure \ref{fig:divide_metapoidataset} assumes that the classification dataset comprises a total of $Y_{K}$ classes. The samples from each class represent a subset of the complete dataset. Based on the attacker's configuration, samples from the $(Y_{K} - 3)$ classes are completely clean, whereas samples from the three targeted classes are categorized into poisoned and clean samples according to the poisoning rate. At this stage, the clean and poisoned subsets are further divided into support sets and query sets, which are subsequently combined to create the poisoned support set and poisoned query set. In the victim model training phase, the meta learner is able to sample from both sets. In Figure \ref{fig:divide_metapoidataset}, the sample sets indicated in green collectively form the poisoned support and query subsets, whereas those marked in yellow denote the corresponding clean support and query subsets. We integrate all poisoned support subsets with the clean support sets to create the poisoned support set and apply the same procedure to the query subsets to form the poisoned query set.

\subsubsection{Meta Learner in PMeta-TLA}

The meta learner manages the interactions between the backdoor meta dataset, the victim classifier, and the weight optimizer, thus controlling gradient propagation among these elements. The process is outlined in Figure \ref{fig:MetaLearner}. In MAML-based meta-learning, the meta-learner functions through two nested loops: (1) selecting training samples from various tasks (Outer loop), and (2) iteratively updating model parameters for each task (Inner loop).

We conceptualize multi-trigger backdoor attacks within the framework of multi-task learning. Applying the meta-learner training procedure of MAML directly results in a proportion of weight updates involving poisoned samples that significantly exceeds the actual poisoning rate, thereby contravening a fundamental principle of backdoor attacks. We present the PMeta-TLA meta-learning training framework to tackle this issue, incorporating the following enhancements:
(1) In the outer loop, tasks are selected so that every $p$ step, a multi-task learning branch is chosen, which includes both backdoor-task training and clean-task training, while the remaining branches adhere to single-task learning, involving only clean-task training. This measure guarantees that, during meta-learning training, the likelihood of the model encountering poisoned samples remains comparable to that in standard backdoor attacks, thus averting a reduction in clean-task accuracy.
(2) The PCGrad algorithm is utilized to calculate weight updates within the multi-task learning framework. Subsequently, we present a comprehensive overview of the meta-learner's workflow.

\noindent\textbf{Step (1). Initialization.} In the initialization step, the victim model $\mathit{C_{\theta}}$ is instantiated along with the specification of the number of triggers $\mathit{TK}$ (i.e., the number of target labels under attack) and the poisoning rate $p\%$. Subsequently, the process proceeds to different branches of Step 2 based on the selection criteria.

\noindent\textbf{Step (2-1). Single-task Learning.} If the current outer-loop epoch index is not equal to $p$, the inner loop, it is assumed to select $\mathit{TK}$ “clean tasks” for training. Accordingly, 
$M \cdot \mathit{TK}$ samples are drawn from the clean support set, and weight updates are performed sequentially on each task. The value $M$ is equivalent to the batch size of the inner loop. This step corresponds to the right-hand branch in Figure \ref{fig:MetaLearner}. The outer-loop learning rate $\lambda_{\alpha}$ is used for weight updates within the outer loop.

\noindent\textbf{Step (2-2). Multi-task Learning.}  If the current outer-loop epoch index is not equal to $P$, the inner loop, it is assumed to select $(\mathit{TK}-1)$ backdoor tasks and one clean task for training. Accordingly, $M$ samples are drawn from the clean support set and $M \cdot (\mathit{TK}-1)$ samples from the poisoned support set. The loss function $L_{S-PT_{i}}$ is computed sequentially for each task, and the model weights are updated using the PCGrad algorithm.  This step corresponds to the left-hand branch in Figure \ref{fig:MetaLearner}.

\noindent\textbf{Step (3). Query Learning.} By this step, the model has already undergone updates across multiple tasks. At this point, the poisoned query set and the clean query set are used to evaluate the model’s performance on backdoor and clean tasks, and the model weights are further updated accordingly with the total sum of loss $L_{Q-T_{j}}$ on each task. Next, depending on the epoch index of the subsequent step, the process re-enters either step (2-1) or step (2-2). The inner-loop learning rate $\lambda_{\beta}$ is used for weight updates within the inner loop.

The proposed meta-learning approach integrates two distinct inner-loop training schemes, alternating between multi-task learning and single-task learning based on the poisoning rate. This guarantees that the frequency of the model's exposure to poisoned samples corresponds with the designated poisoning rate.

\subsection{Robustness Against Potential Defenses}

Many backdoor defenses have been proposed to mitigate threats in image classification \cite{guo2023scale,xiang2023umd,jebreel2023defending}, yet most are not directly applicable to speech classification tasks due to their image-specific design. Here, we evaluate the proposed attack against five representative cross-domain defenses: fine-tuning \cite{liu2017neural}, pruning \cite{liu2018fine}, STRIP \cite{gao2019strip}, spectral signatures \cite{tran2018spectral}, and trigger filtering \cite{du2019robust}. We train each victim model on the KWS dataset to evaluate the capability of the PMeta-TLA method in resisting these defense mechanisms. We will describe the specific procedures for resisting various defense methods in the experimental setup section.

\section{Experiments and Results}
\label{sec5}

\subsection{Experimental Setting}

\noindent\textbf{Dataset.} We assess our approach on keyword spotting tasks. We utilized Google Speech Commands version 2 (GSCv2) for KWS \cite{speechcommandsv2}. The GSCv2 comprises 105,829 utterances associated with 35 keywords. Each audio file representing an individual utterance has a duration of approximately one second, with the active speech command comprising over fifty percent of this total duration. The dataset configuration was based on the VSVC \cite{cai2022vsvc} paper, resulting in the extraction of 10 keyword classes, comprising a total of 65,000 audio files, to enable a comparison of attack effectiveness. A total of 1500 audio files were created by randomly selecting 150 samples from each class, forming a poisoned test set used to assess the effectiveness of the attack following backdoor training.

\noindent\textbf{Baseline Backdoor Attacks.}
We compare our attack with the latest speech backdoor attacks. They are as follows: (1) Position-independent backdoor attack (PIBA) \cite{shi2022audio}, (2) Dual-adaptive backdoor attack (DABA) \cite{liu2022opportunistic}, (3) Ultrasonic voice as the trigger (Ultrasonic) \cite{koffas2022can}, (4) Pitch boosting and sound masking (PBSM) \cite{cai2022pbsm}, and (5) Voiceprint selection and voice conversion (VSVC) \cite{cai2022vsvc}.

\begin{table*}[ht]
\centering
\caption{Attack results on the GSCv2 dataset towards the KWS task. Each item shows the ASR (\%) / PN in the table.}
\begin{tabularx}{0.8\textwidth}{l|*{4}{>{\centering\arraybackslash}X}}
\toprule
  ASR ($\uparrow$)  /  PN ($\downarrow$)         & ERes2Net & KWS-ViT & EAT-S & CAM++ \\ 
\midrule
PIBA \cite{shi2022audio}       & 95.33 / 550 & 96.46 / 500 & 95.93 / 550  & 94.80 / 600 \\
DABA \cite{liu2022opportunistic}       & 94.26 / 450 & 93.33 / 450 & 92.13 / 500 & 92.53 / 500 \\
Ultrasonic \cite{koffas2022can} & 95.40 / 400 & 94.93 / 450 & 93.87 / 450 &  93.53 / 500 \\
PBSM \cite{cai2022pbsm}       & 97.13 / 350 & 98.87 / 400 & 98.93 / 450 &  98.20 / 450 \\
VSVC \cite{cai2022vsvc}       & 99.13 / 300 & 99.27 / 350 & 98.53 / 350 & 97.27 / 400 \\ 
\midrule

TLA-S (t=1) & 98.93 / 350 & 99.13 / 400 & 98.60 / 400 & 97.47 / 450\\ 
TLA-M (t=3) & 98.80 / $(400\times3)$ & 98.47 / $(450\times3)$ & 98.20 / $(450\times3)$ & 97.13 / $(450\times3)$ \\ 
PMeta-TLA (t=3) & 99.67 / $(300\times3)$ & 99.40 / $(320\times3)$ & 99.13 / $(350\times3)$ & 98.20 / $(450\times3)$ \\ 
PMeta-S (t=3+1) & 98.53 / 250 & 98.337 / 260 & 97.60 / 260 & 97.07 / 280 \\ 

PMeta-M (t=3+3) & 97.20 / $(250\times3)$ & 97.60 / $(260\times3)$ & 96.40 / $(260\times3)$ & 95.93 / $(250\times3)$ \\
\bottomrule

\end{tabularx}
\label{table:kws_res}
\end{table*}

\begin{figure*}[ht]
    \centering
    \includegraphics[width=\linewidth]{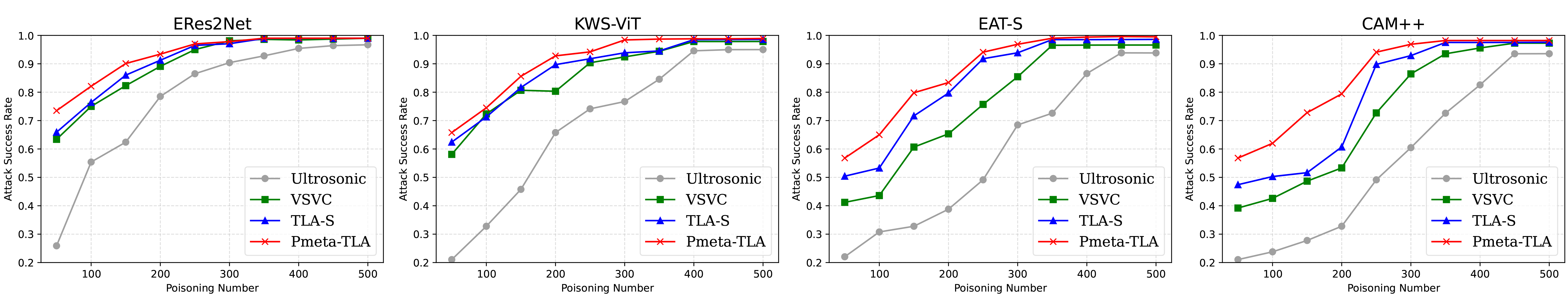}
     \caption{The figure illustrates the ASR-PN curves for various backdoor-attack methods evaluated on three baseline KWS models. Among them, Ultrasonic and VSVC serve as the baseline approaches, whereas TLA-S and Pmeta-TLA represent the proposed novel methods.}
    \label{fig:c1}
\end{figure*}

\begin{table*}[tbp]
\centering
\caption{ASR (\%) / PN results when five types of backdoors (t=5) are simultaneously implanted during training.}
\begin{tabular}{lcccc}
\hline
 & ERes2Net & KWS-ViT & EAT-S & CAM++ \\
\hline
TLA-S (t=1)        & 98.20 / 350 & 97.33 / 400 & 97.12 / 400 & 97.83 / 400 \\
TLA-M (t=5)              & 96.89 / 400 & 97.52 / 400 & 96.73 / 450 & 96.87 / 450 \\
TLA-M (w/ meta)    & 97.92 / 300 & 98.92 / 300 & 97.40 / 350 & 97.07 / 350 \\
TLA-M (w/ PCGrad)  & 98.80 / 400 & 99.47 / 450 & 99.20 / 450 & 99.07 / 450 \\
Pmeta-TLA          & 98.93 / 300 & 99.67 / 350 & 99.40 / 350 & 98.93 / 350 \\
\hline
\end{tabular}
\label{table:t3}
\end{table*}

\noindent\textbf{Victim Model.} Our experiments were performed on the KWS and SV tasks. For the KWS task, we used the latest SOTA models, like ERes2Net \cite{chen2023enhanced}, KWS-ViT \cite{berg2021keyword}, EAT-S \cite{gazneli2022end}, CAM++ \cite{wang2023cam++}, which exhibit excellent classification performance on the keyword spotting task. It should be noted that the ERes2Net and CAM++ originally designed for lightweight speaker verification, were repurposed for the KWS task by modifying their last weight layers and loss functions.

\noindent\textbf{Backdoor Training Protocols.} To systematically evaluate the efficacy of the TLA trigger and the meta-learning paradigm, we curate a set of out-of-distribution timbres from an auxiliary speech corpus and design five complementary experimental protocols. In the backdoor attack experiments, $t$ denotes the number of backdoors injected. \textbf{(1) TLA-S.
(t=1)} Single-target backdoor attack that employs a single timbre as the TLA trigger. \textbf{(2) TLA-M (t=3).} Multi-target backdoor attack that leverages multiple timbres as TLA triggers. In this experiment, we used three timbres as triggers. \textbf{(3) PMeta-TLA (t=3).} Multi-target backdoor attack that integrates multiple timbres as TLA triggers that trained with the proposed PMeta-TLA framework. In this experiment, we used three timbres. In addition, to probe the “learn-to-learn” capability inherent in the meta-learning formulation, we devise two further protocols:  \textbf{(4) PMeta-S (t=3+1).} On the basis of Pmeta-TLA (t=3), this experimental configuration implants a new backdoor into the model during the meta-fine-tuning stage (This is the meaning of “+1”), and \textbf{(5) PMeta-M (t=3+3).} Building upon the three backdoors already embedded by Pmeta-TLA, this experimental configuration is designed to inject three additional backdoors incrementally, resulting in a victim model that ultimately harbors six backdoors. In Experiment PMeta-M, the model undergoes meta fine-tuning initialized from the victim model obtained in Experiment PMeta-TLA. In this experiment, we additionally used three timbre triggers for meta fine-tuning (this is the meaning of “+3”). 

According to our proposed method, meta-learning backdoor attacks enable the model to learn how to implant backdoors and, given only a few trigger–target-label pairs, rapidly acquire the ability to inject a new backdoor. Therefore, by designing experiments that implant one or three novel backdoors during the fine-tuning phase, we seek to empirically validate the effectiveness of the proposed approach.

\noindent\textbf{Training Setup.}
The batch size for the KWS task is 64. The weights are optimized using the Adam optimizer \cite{kingma2015adam} with a cross-entropy loss function. We conducted training for 40 epochs to ensure the convergence of all models. Throughout the meta-training and meta-finetuning phases, the optimizer utilizes the same learning rate and hyperparameters. The learning rate is established $\lambda_{\alpha}=2e-4$ for support-set training and $\lambda_{\beta}=1e-4$ for query-set training.

\subsection{Evaluation Metrics.}
The evaluation metrics include attack metrics and a trigger metric. The attack metrics include ASR, and poisoning number (PN, which equals the number of $D_{sp}$). 

\noindent\textbf{ASR.} The Attack Success Rate (ASR) is defined as the percentage of poisoned test samples that the victim model misclassifies as the target label. An effective attack is characterized by a high ASR.

\noindent\textbf{PN.} The Poisoned number (PN) intuitively indicates the quantity of poisoned samples utilized by each approach to get the best ASR. The frequently utilized backdoor metric, the "poisoning rate" (the ratio of poisoned samples to the total number of test samples), is generally minimal in relation to the overall dataset size and varies across datasets; thus, we deem this proportional measure inappropriate for comparative analysis. We utilize the quantity of poisoned samples as the evaluation parameter, which enhances comparability when assessing variations in attack efficacy across various models on the same dataset. When PN is low and ASR is high, it indicates that the trigger exhibits robust attack efficacy.

\begin{figure}[t]
    \centering
    \includegraphics[width=\linewidth]{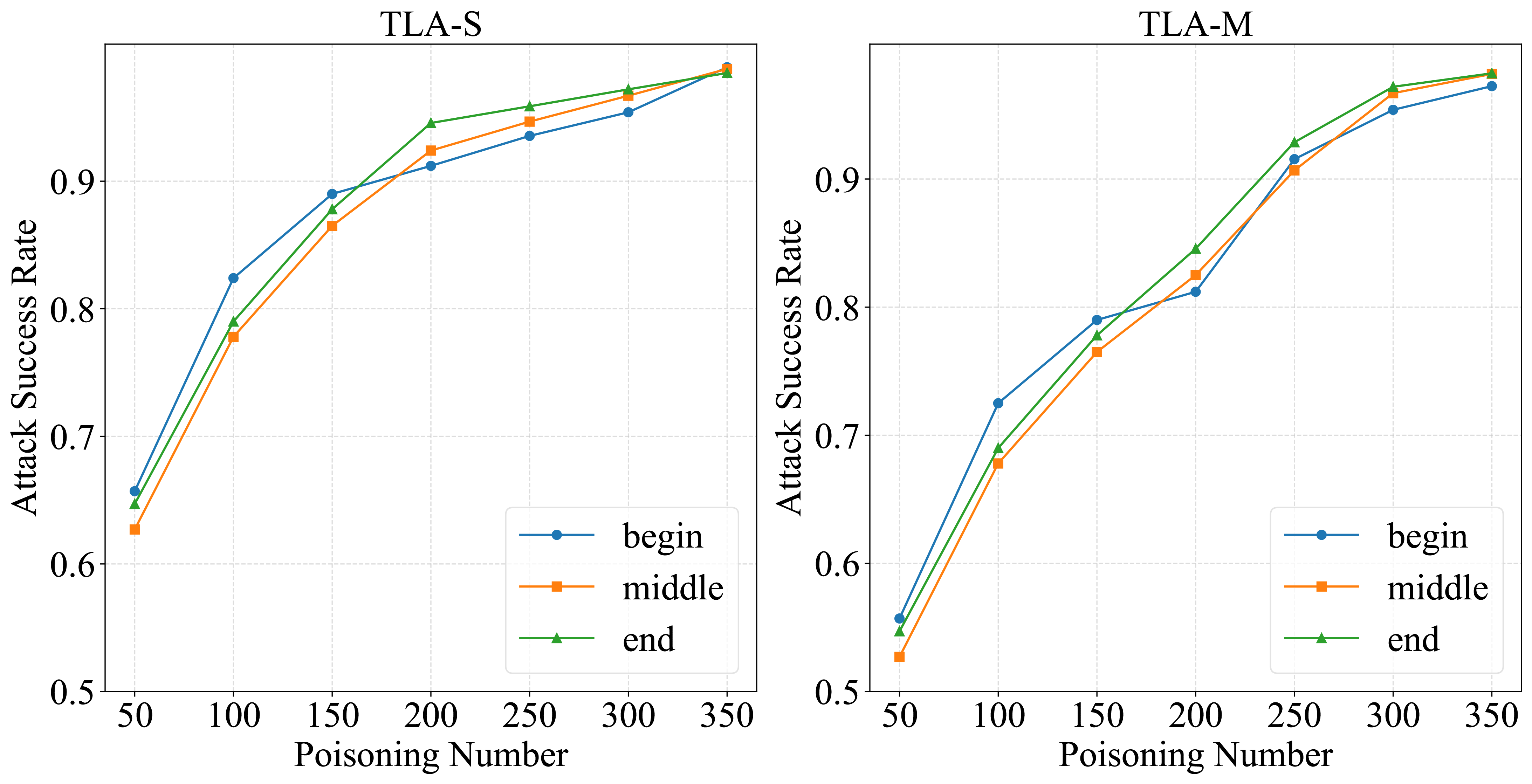}
     \caption{The figure presents the attack success rates of the victim model on the poisoned test set, evaluated under proposed TLA triggers with different leakage positions, as the number of poisoning number increases.}
    \label{fig:c2}
\end{figure}

\subsection{Ablation Study}

In the validation experiments of timbre-leakage triggers, we further considered the effects of leakage position and target timbre on the attack performance.

\noindent\textbf{Leakage Position.} Within the trigger function TLA, a designed algorithm replaces a short segment of the speech signal with the target timbre while preserving semantic content and other speech attributes. The choice of segment position may affect the trigger’s effectiveness. We set the timbre-leakage position to the beginning, middle, or end of the speech segment. During attacks, poisoned samples may contain the leakage at a random position or with the leakage fixed at a specific position. The beginning position refers to the first set of discrete pronunciation units, while the end position corresponds to the last set of discrete pronunciation units. The middle position denotes any location between these two. We show the experiment results on Figure \ref{fig:c2}.

To evaluate whether there were significant differences among the ASR under the three leakage position conditions, we performed a one-way Analysis of Variance (ANOVA). The analysis yielded an F-statistic of $F=0.01775$ and a p-value of $p=0.9824$. As the p-value is substantially greater than the conventional significance level of $\alpha=0.05$, we did not find any statistically significant difference between the means of the three groups. Therefore, we conclude that the leakage location does not have a significant impact on the attacks.

\subsection{Ablation Study}

In the validation studies of timbre-leakage triggers, we additionally examined the influence of leakage position and target timbre on backdoor attack performance.

\noindent\textbf{Leakage Position.} In the trigger function TLA, a specialized algorithm substitutes a brief section of the voice signal with the desired timbre while maintaining semantic content and other speech characteristics. The selection of segment position may influence the trigger's efficacy. We configured the timbre-leakage position to the beginning, middle, or end of the speech segment. During our backdoor attack, poisoned samples may exhibit leakage at a random location or with leakage confined to a designated position. The beginning position denotes the initial set of discrete pronunciation units, whereas the end position signifies the final set of discrete pronunciation units. The central position signifies any location situated between these two points. The experimental results are presented in Figure \ref{fig:c2}.

We used a one-way Analysis of Variance (ANOVA) to assess significant differences in ASR across the three leaking position circumstances. The study produced an F-statistic of $F=0.01775$ and a p-value of $p=0.9824$. The p-value, being significantly higher than the standard significance threshold of $\alpha=0.05$, indicates that there is no statistically significant difference among the means of the three groups. Consequently, we ascertain that the leakage site exerts minimal influence on the attacks.

\noindent\textbf{Multi-backdoor Strategy.}
Our proposed methodology utilises a meta-learning strategy to mitigate the poisoning cost associated with backdoor attacks and incorporates the PCGrad algorithm to augment attack efficacy. These strategies are pertinent to multi-backdoor injection attacks. We also examine the effects of various tactics on a baseline TLA backdoor attack, as seen in Table \ref{table:t3}. We developed a single-target backdoor attack and five-target backdoor attacks augmented by four distinct strategies. The data in Table \ref{table:t3} indicates that the PN necessary for multi-target backdoor attacks exceeds that required for single-target attacks, but the ASR is marginally reduced. Implementing the meta-learning technique results in a significant decrease in PN, around 100 samples. The PCGrad approach yields a slight enhancement in ASR (about $1\%$). When both tactics are utilised together, the backdoor attacks attain a greater Attack Success Rate (ASR) with a reduced number of poisoned samples.

In conclusion, we found that the meta-learning strategy enables the model to learn discriminative features between poisoned and clean speech samples across different backdoor tasks, thereby reducing the poisoning number required for each task. In contrast, the PCGrad strategy mitigates conflicting gradients between backdoor tasks and clean tasks, which enhances the accuracy of each task and consequently improves the overall ASR.

\subsection{Defense Setting and Results}

To verify the robustness of our proposed timbre-leakage trigger against five defense methods, we conducted defense-resistance experiments on two backdoored models with different architectures, CAM and EAT, which were poisoned using the KWS dataset and our proposed Pmeta-TLA method.

\noindent\textbf{Fine-tuning Defense.} As a typical backdoor removal method, fine-tuning \cite{liu2017neural} aims to eliminate model backdoors by using a small set of local benign samples to fine-tune the model. The motivation for this approach stems from the catastrophic forgetting property of deep neural networks \cite{kirkpatrick2017overcoming}. We use $10\%$ of the benign data as training samples to fine-tune the victim model.

\begin{figure}[h]
    \centering
    \includegraphics[width=\linewidth]{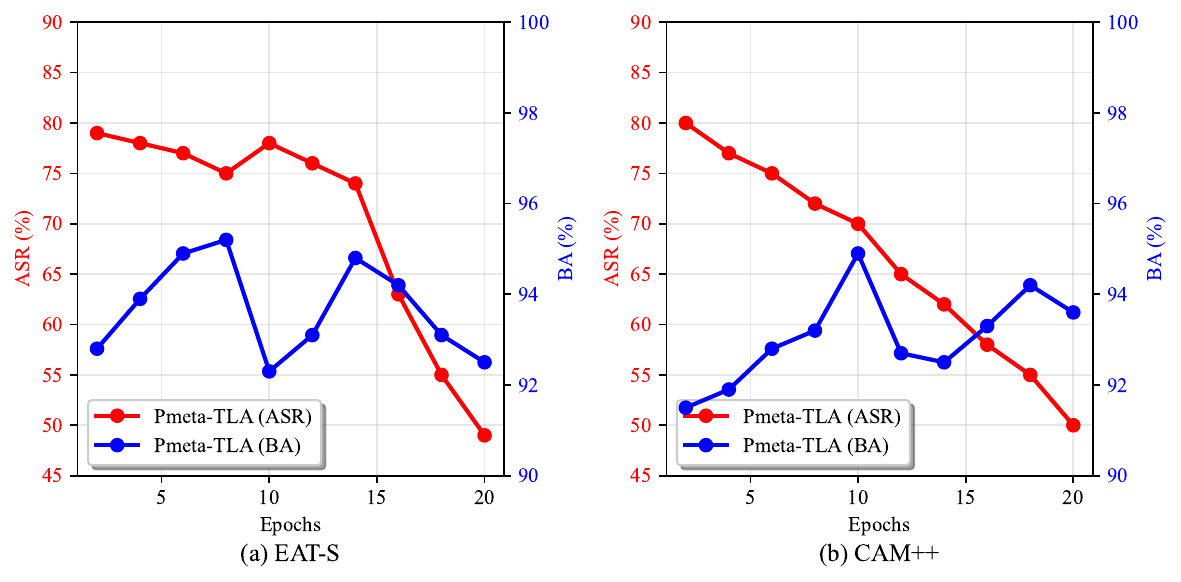}
    \caption{Pmeta-TLA Performance under the model fine-tuning defense.}
    \label{fig:defen-finetune}
\end{figure}

As shown in the Figure \ref{fig:defen-finetune}, the attack success rate decreases with the increase of fine-tuning epochs. However, even at the end of this process, the attack success rate of Pmeta-TLA remains above 45\% on both models. These results demonstrate that the proposed attack can largely withstand fine-tuning.

\noindent\textbf{Pruning Defense.}  Model pruning \cite{liu2018fine} aims to remove model backdoors by pruning neurons that remain inactive during the inference of benign samples. The motivation behind this method lies in the assumption that, in a backdoored DNN, the backdoor-related neurons and benign neurons are mostly separate. When a large number of neurons are pruned, the attack success rate drops significantly; however, this comes at the cost of a sharp decline in benign accuracy.

\begin{figure}[h]
    \centering
    \includegraphics[width=\linewidth]{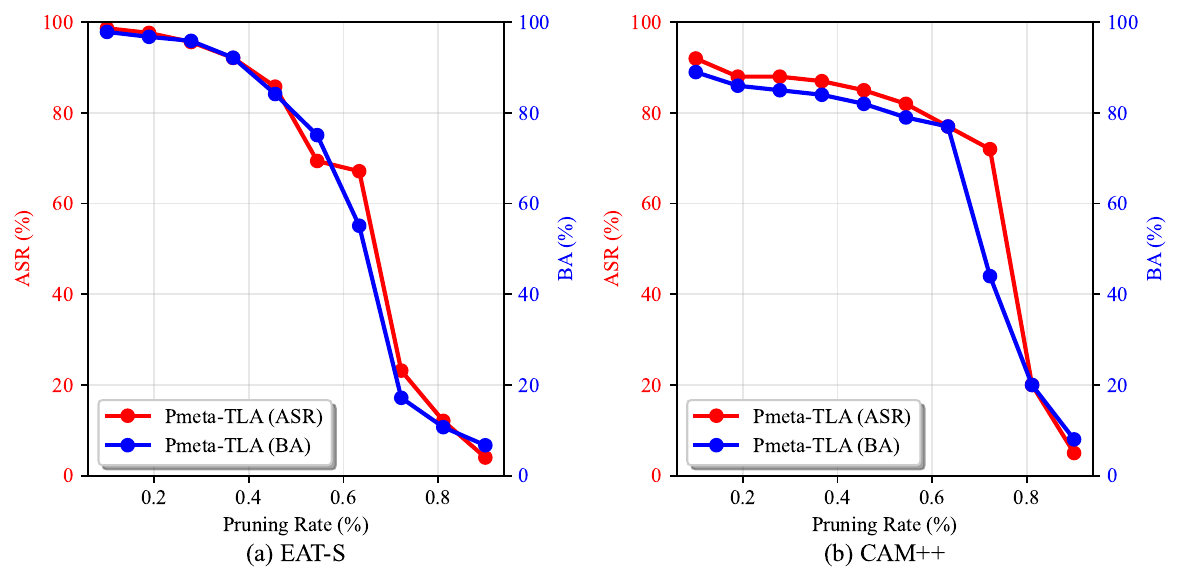}
    \caption{Pmeta-TLA Performance under the model pruning defense.}
    \label{fig:defen-prune}
\end{figure}

The Figure \ref{fig:defen-prune} illustrates that a substantial reduction in the number of neurons leads to a large decrease in the attack success rate. Nonetheless, this results in a significant reduction in benign accuracy. For Pmeta-TLA, the reduction in attack success rate closely parallels the decline in benign accuracy. The primary reason is that the assumption foundational to model pruning is invalid for Pmeta-TLA attacks, as their triggers are crafted to be both global and intricate. The results illustrate the robust resilience of the Pmeta-TLA attack against model pruning.

\noindent\textbf{STRIP Defense.} As a representative black-box technique that detects poisoned samples by analyzing predicted logit values, STRIP \cite{gao2019strip} introduces perturbations by superimposing test samples with various other samples and then examines the entropy of the model’s predictions. Samples with low entropy are regarded as suspicious and classified as poisoned. The robustness of Pmeta-TLA against STRIP is evaluated by visualizing the entropy distribution of the samples.

\begin{figure}[h]
    \centering
    \includegraphics[width=\linewidth]{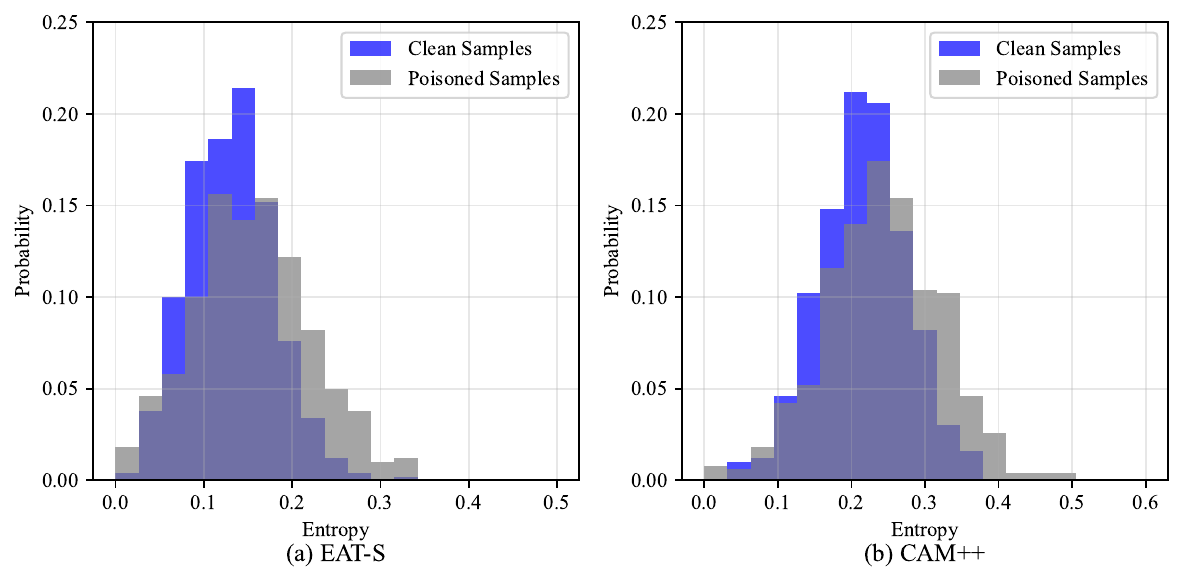}
    \caption{Pmeta-TLA Performance under the model pruning defense.}
    \label{fig:defen-STRIP}
\end{figure}

As shown in the Figure \ref{fig:defen-STRIP}, the clean and poisoned samples exhibit similar and inseparable entropy distributions in both models. Therefore, Pmeta-TLA can resist detection by STRIP.

\begin{figure}[h]
    \centering
    \includegraphics[width=\linewidth]{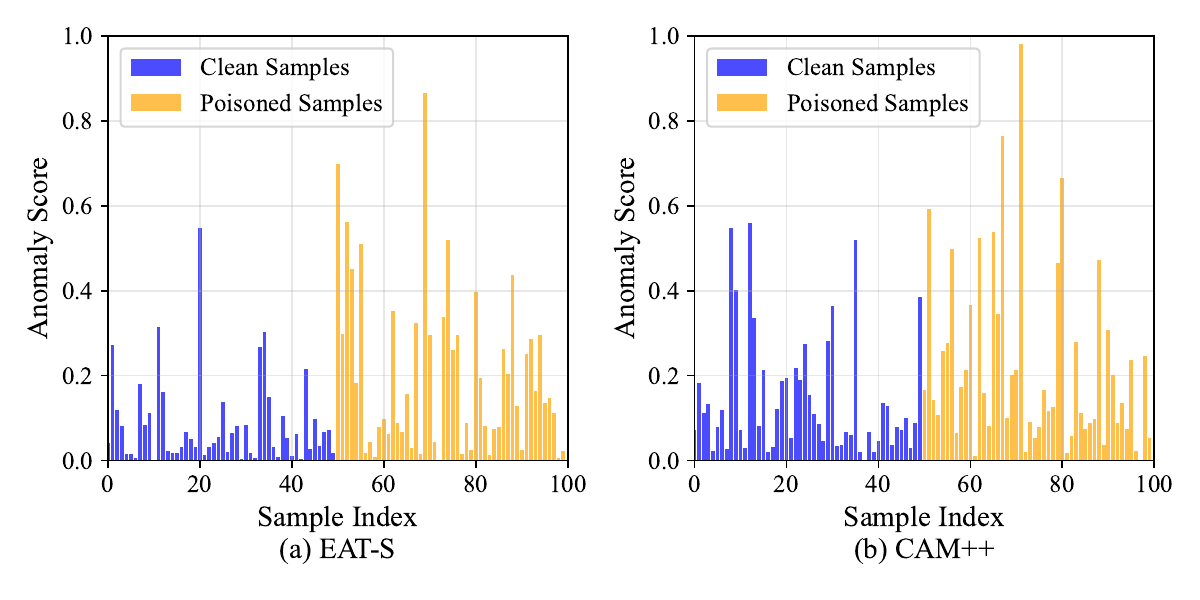}
    \caption{Pmeta-TLA Performance under the spectral signature defense.}
    \label{fig:defen-spectral}
\end{figure}

\noindent\textbf{Spectral Signatures  Defense.} The spectral signature defense method \cite{tran2018spectral} suggests that in the high-level feature representations of neural networks, backdoor signals are significantly amplified, causing the feature representations of poisoned samples to exhibit a notable statistical shift—particularly in the mean—relative to those of clean samples under the same label. The spectral signature method uses the first principal component obtained through Singular Value Decomposition (SVD) to compute an anomaly score for each sample, where poisoned samples typically have higher scores and are removed from the poisoned training set. As shown in the Figure \ref{fig:defen-spectral}, we calculated the distribution of anomaly scores for two victim models. We tested 100 samples in total, with samples 0–49 being clean and samples 50-100 being poisoned.

The anomaly scores of the poisoned samples produced by the Pmeta-TLA approach are often marginally higher than those of the clean samples; however, the overall discrepancy in anomaly values is minimal. This suggests that the approach can withstand the spectral signature defense to some degree.

\noindent\textbf{Trigger Filtering  Defense.} To deactivate potential backdoors in a compromised DNN, defenders may remove high-frequency signals, low-frequency signals, and noise to eliminate potential trigger patterns in suspicious test speech.

As shown in the Table \ref{table:denfen-trigger}, we incorporated three types of trigger filtering methods—high-frequency signal removal, low-frequency signal removal, and the addition of babble noise—along with their combinations when calculating the ASR, to observe the degree of ASR degradation. We found that the ASR of both TLA and Pmeta-TLA remained almost unchanged, demonstrating the strong resistance of the timbre-leakage trigger against this defense.

We found that several trigger filtering methods failed to effectively reduce the ASR of backdoor attacks, demonstrating the resistance capability of Pmeta-TLA.

\begin{table}[h]
\centering
\caption{The ASR of victim models on the backdoor test set using the trigger filtering defense method.}
\begin{tabular}{c|cc}
\hline
Trigger Filter                & EAT-S & CAM++ \\ \hline
high-frequency signal removal & 98.27 & 97.73 \\
low-frequency signal removal  & 98.60 & 98.07 \\
babble noise                  & 96.20 & 97.13 \\
All                           & 94.93 & 95.67 \\ \hline
\end{tabular}
\label{table:denfen-trigger}
\end{table}

\section{Conclusion}
\label{sec6}

In this paper, we propose Pmeta-TLA, a system-level backdoor attack for speech models that incorporates an innovative timbre-leakage attack function (TLA) together with backdoor training and fine-tuning procedures that leverage meta-learning and PCGrad. We formalize the backdoor attack as a joint training problem involving both clean and backdoor tasks, employ PCGrad to mitigate gradient conflicts between these tasks, and adopt a meta-learning strategy to teach the model to acquire the ‘implantation’ behavior. Experimental results on KWS task demonstrate that our method outperforms conventional single-target backdoor attacks. Moreover, Pmeta-TLA is capable of implanting multiple backdoors in a single training run, substantially increasing the difficulty for defenders to detect or remove the malicious behavior. We anticipate that further exploration of speech backdoor techniques will contribute valuable insights for the design of more robust defenses.

\bibliographystyle{elsarticle-num}
\bibliography{manuscript}

\end{document}